\documentclass[aps,reprint,superscriptaddress,nofootinbib,preprintnumbers,longbibliography]{revtex4-2}

\usepackage{url}
\usepackage{upgreek,amssymb,multirow}
\usepackage{mathtools}
\usepackage{float}
\usepackage{mciteplus}
\usepackage{blindtext}

 \usepackage{dsfont}

\usepackage{comment}

\usepackage{amsmath}
\usepackage{extarrows}
\usepackage{braket}

\usepackage[dvipsnames,svgnames]{xcolor}

\usepackage{tikz-cd}
\usetikzlibrary{calc,topaths,decorations,decorations.pathmorphing,arrows,decorations.markings,cd}
\tikzset{
->-/.style args={#1rotate#2}{decoration={markings, mark=at position #1 with {\arrow[scale=1.5,rotate = #2 ]{stealth}}}, postaction={decorate}}
}

\tikzset{line/.style={line width=0.25mm},
curve/.style={line,smooth,tension=1},
->-/.style={decoration={
  markings,
  mark=at position #1 with {\arrow[>=stealth]{>}}},postaction={decorate}},
-<-/.style={decoration={
  markings,
  mark=at position #1 with {\arrow[>=stealth]{<}}},postaction={decorate}},
}

\usepackage{breakurl}
\usepackage[hyperindex,breaklinks]{hyperref}

\newcommand{\ie}{\begin{equation}\begin{aligned}}
\newcommand{\fe}{\end{aligned}\end{equation}}

\newcommand{\D}{\mathsf{D}}

\newcommand{\cN}{{\mathcal{N}}}
\newcommand{\cD}{{\mathcal{D}}}

\newcommand{\cH}{{\cal H}}

\newcommand{\cC}{{\cal C}}

\newcommand{\Tr}{\mathrm{Tr}}

\newcommand{\ra}{\rightarrow}
\newcommand{\Hom}{{\rm Hom}}

\newcommand{\Rep}{{\rm Rep}}

\newcommand{\CC}{{\mathbb C}}
\newcommand{\ZZ}{{\mathbb Z}}

\newcommand{\cE}{{\mathcal E}}

\newcommand{\cA}{\mathcal A}
\newcommand{\cL}{\mathcal L}
\newcommand{\cO}{\mathcal O}

\begin{document}

\title{Symmetric entanglers for non-invertible symmetry protected topological phases}
 
\author{Minyoung You}
\email{miyou849@gmail.com}
\affiliation{ Yukawa Institute for Theoretical Physics, Kyoto University, Kitashirakawa Oiwakecho,
 Sakyo-ku, Kyoto 606-8502, Japan}

\begin{abstract}
It has been suggested that non-invertible symmetry protected topological phases (SPT), due to the lack of a stacking structure, do not have symmetric entanglers (globally symmetric finite-depth quantum circuits) connecting them. Using  topological holography, we argue that a symmetric entangler should in fact exist for 1+1d systems whenever the non-invertible symmetry has SPT phases  connected by fixed-charge dualities (FCD). Moreover, we construct an explicit example of a symmetric entangler for the two SPT phases with $\mathrm{Rep}(A_4)$ symmetry, as a matrix product unitary (MPU).
\end{abstract}


\maketitle

\section{Introduction and background}

A fundamental property of symmetry protected topological (SPT) phases is that they cannot be distinguished from each other in the bulk -- the difference between the phases is only visible at the edge. This is ensured by the fact that they are connected by symmetric entanglers -- globally (but not locally) symmetric finite-depth quantum circuits (FDQC) \cite{chen2012symmetry, Chen_2013, zeng2018quantuminformationmeetsquantum, Potter_2018, tantivasadakarn2023pivot} -- which map local operators to local operators while preserving their charges under the symmetry. Symmetric entanglers also encode important properties of SPT phases,  such as the SPT invariants and the group law under the stacking of  phases \cite{Ellison_2019,zhang2023topological}.  

The best-known example is the entangler for the  1d  cluster model, with the Hamiltonian 
$$H_{\rm cluster} = - \sum_j  Z_{j-1}  X_j Z_{j+1}$$
where $Z_j, X_j$ are Pauli operators acting on qubits at site $j$. This belongs to the nontrivial SPT phase protected by $\ZZ_2\times \ZZ_2$ symmetry generated by $\prod \limits_{j \text{ odd} } X_j$ and $\prod \limits_{j \text{ even} } X_j$.
If we denote its ground state by $|\Psi_{\rm cluster}\rangle$, we can obtain this via acting with the entangler 
$$\cE = \prod_{j} \mathrm{CZ}_{j,j+1}$$
on the product state, as
$$|\Psi_{\rm cluster}\rangle = \cE |\Psi_{\rm prod}\rangle.$$ 
Here, ${\rm CZ}_{j, j+1}$ are the controlled-$Z$ gates on sites $j, j+1$. Note that the entangler $\cE$ commutes with the global $\ZZ_2\times\ZZ_2$ symmetry generators, even though its local pieces ${\rm CZ}_{j,j+1}$ do not.  
We also note that $$\cE^2 = \mathds{1};$$
hence, applications of $\cE$ play the role of stacking with the SPT state, and reproduce the group structure  $H^2(\ZZ_2 \times \ZZ_2, U(1)) = \ZZ_2$ coming from group cohomology  \cite{seifnashri2024cluster}. The relation between such entanglers and stacking holds in general for ordinary symmetries, as discussed in Sec. IV. A of Ref. \cite{Ellison_2019}.

In recent years, SPT phases protected by non-invertible symmetries (non-invertible SPT phases), which in 1+1d form a fusion category rather than a group, have received much attention \cite{thorngren2024fusion,zhang2024anomalies, Bhardwaj_2025, Inamura_2021, inamura202411dsptphasesfusion, Fechisin_2025, seifnashri2024cluster, meng2025noninvertiblesptsonsiterealization}. Here, the story of symmetric entanglers gets more complicated: Ref. \cite{seifnashri2024cluster}, which studied models realizing $\Rep(D_8)$ SPT phases in detail, showed that there is no symmetric entangler connecting them. Moreover, they suggested that this is a universal fact about non-invertible SPT phases, originating from their lack of a stacking structure (see Refs. \cite{Chen_2013,Bultinck_2017,Kapustin_2018,Turzillo_2019, ren2024stacking, Turzillo_2024} for the stacking structure of SPT phases).

This is in tension with Ref. \cite{aksoy2025}, which proposed a classification framework for non-invertible SPT phases in which phases related by a class of dualities which preserve charges (fixed-charge dualities, or FCDs) belong to an \emph{SPT class}. Their proposal is based on the fact that, in the case of ordinary symmetries, symmetric entanglers implement the operation of stacking by an SPT phase \cite{moradi2023topological, Turzillo_2019, kobayashi2025soft, kobayashi2025projective,aksoy2025}, which is a special case of FCDs. If we return to the cluster model example, the symTFT is given by two copies of the Toric Code MTC (with its group of anyons generated by $e_1, m_1, e_2, m_2$), and the duality which maps 
$$m_1 \mapsto m_1 e_2, \quad m_2 \mapsto m_2 e_1$$
(while preserving $e_1, e_2$)
is an FCD which maps the trivial phase to the SPT phase (and vice versa -- thus, implementing  ``stacking with the SPT phase'').
This suggests an intimate connection between FCDs and symmetric entanglers, which led Ref. \cite{aksoy2025} to their classification proposal. 
However, they did not construct the FCDs on the lattice, and thus their proposal lacked a microscopic foundation.

 We resolve this problem by showing that for 1+1d non-invertible SPT phases, symmetric entanglers can in fact exist. There are two main results, which are complementary and self-contained. 
In Sec. \ref{sec:FCD}, we prove on the level of topological holography/fusion categories that when a duality preserves the symmetries of the 1+1d system if and only if it is an FCD. Based on this mathematical result,  we conjecture  that an FCD leads to a symmetric entangler for SPT phases.  Sec. \ref{sec:A4} is less general but more concrete: we  explicitly construct a symmetric entangler as a matrix product unitary (MPU) connecting the two SPT phases of $\Rep(A_4)$. 
 Taken together with the result of Ref. \cite{seifnashri2024cluster} that some non-invertible SPT phases do not have symmetric entanglers connecting them, this means that non-invertible SPT phases are not all on an equal footing -- some phases are \emph{more different} from each other than others. While the exact relationship between FCDs and symmetric entanglers remain conjectural, this example provides evidence that there is a microscopic basis for the classification of non-invertible SPT phases into SPT classes  proposed  in Ref. \cite{aksoy2025}.

\section{Fixed-charge dualities preserve symmetries}\label{sec:FCD}

\begin{figure}[t]
    \centering
\raisebox{-73pt}{\begin{tikzpicture}
\draw [<->] (1,4.2) -- (3,4.2);
\draw (2,4.2) node[above]{$I$};
\draw (1.35,0.8) node[below]{\small \color{red!75!DarkGreen} $Z(\cC)$};
\draw[color=DarkGreen] (0,0) -- (0,3) -- (1,4) -- (1,1) -- cycle;
\draw (0,0) node[below]{\color{DarkGreen} $\mathbb{B}_Q$};
\draw [color=red!75!DarkGreen, thick, decoration = {markings, mark=at position 0.6 with {\arrow[scale=1]{stealth}}}, postaction=decorate] (0.5,2) -- (1.5,2) node[below]{$\mu$} -- (2.5,2);
\draw [fill=DarkGreen] (0.5,2) circle (0.04) node [below] {\color{DarkGreen} $V_\mu$};
\draw [color=blue!70!green, thick, decoration = {markings, mark=at position 0.5 with {\arrow[scale=1]{stealth}}}, postaction=decorate] (2.5,2) -- (2.5,3) node[right]{$a$} -- (2.5,3.75);
\draw [fill=blue!70!green] (2.5,2) circle (0.04) node [below] {\color{blue!70!green} $W_a^\mu$};
\draw[color=blue!70!green, preaction={draw=white,line width=3pt}] (2,0) -- (2,3);
\draw[color=blue!70!green] (2,3) -- (3,4) -- (3,1) -- (2,0);
\draw (2,0) node[below]{\color{blue!70!green} $\D$};
\end{tikzpicture}}
    \quad $=$ ~
\raisebox{-73pt}{\begin{tikzpicture}
\draw [color=blue!70!green, thick, decoration = {markings, mark=at position 0.5 with {\arrow[scale=1]{stealth}}}, postaction=decorate] (2.5,2) -- (2.5,3) node[right]{$a$} -- (2.5,3.75);
\draw [fill=black] (2.5,2) circle (0.04) node [below] {$\cO$};
\draw(2,0) -- (2,3) -- (3,4) -- (3,1) -- cycle;
\draw (2,0) node[below]{$Q$};
\end{tikzpicture}}
    \caption{SymTFT setup on $\Sigma \times I$, where $\Sigma$ is a $2$-manifold and $I$ is the interval. $\D$ is the reference Dirichlet boundary condition, and $\mathbb{B}_Q$ is the physical boundary condition. $V_\mu$ is the space of local operators which tells us how  the bulk anyons $\mu$ can end on the physical boundary.  $W_a^\mu$ is the space of junctions between $\mu$ and the symmetry line $a$ which lives on the reference boundary. Compactifying the interval leads to a $\cC$-symmetric 1+1d system $Q$, with the anyon $\mu$ turning into an $a$-twisted sector local operator $O$. 
    \cite{Lin_2023}
    }
    \label{fig:symTFT}
\end{figure}
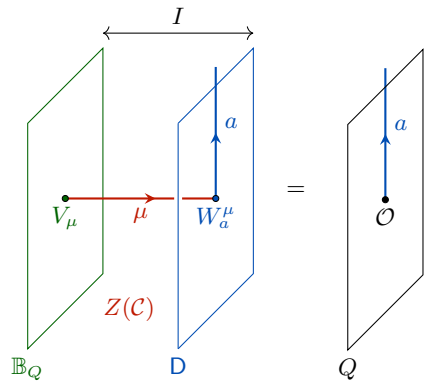

The fusion category symmetry $\cC$ of a 1+1d system may be interpreted in terms of topological line  operators (anyons) of a 2+1d topological quantum field theory (TQFT), which is mathematically described by a modular tensor category (MTC) given by the Drinfeld center $Z(\cC)$ of $\cC$ -- this framework is known as topological holography, and the 2+1d TQFT is referred to as ``symTFT'' in this context \cite{Gaiotto_2021,moradi2023topological, Lin_2023,bhardwaj2023generalized,  zhang2024anomalies,Bhardwaj_2025, choi2024generalizedtubealgebrassymmetryresolved}. See Figure \ref{fig:symTFT} for the setup. 
We impose the (reference) Dirichlet boundary condition $\D$ and a physical boundary condition $\mathbb{B}_Q$ at the two boundaries.
There is a forgetful functor 
$$F: Z(\cC) \ra \cC,$$
from the bulk to the reference boundary, which sends a bulk line operator (anyons of the TQFT) $\mu$ to a boundary line operator  $F(\mu)$ (which is a symmetry operator of the 1+1d theory), which may be non-simple \cite{putrov2024nonanomalousnoninvertiblesymmetries11d}.  

The 
$a$-twisted sector Hilbert space decomposes as 
$$\cH_a= \bigoplus_\mu W_a^\mu \otimes V_\mu,$$
where $W_a^\mu$ is the space of junctions between the bulk line $\mu$ and the boundary line $a$, and $V_\mu$ is the space of local operators on the physical boundary where $\mu$ can end \cite{Lin_2023, choi2024generalizedtubealgebrassymmetryresolved}.
We define 
$$N_a^\mu := \dim W_a^\mu = \langle F(\mu), a \rangle_{\cC} \equiv  \Hom_{\cC}(F(\mu), a).$$ 
Then, 
\begin{align}
    F(\mu) = \bigoplus \limits_{a\in {\rm Irr}(\cC)} N_a^\mu a
    \label{eq:F}
    \end{align}
tells us how the anyon $\mu$ transmutes into boundary symmetry lines.

A boundary condition is specified by a Lagrangian algebra of anyons condensed on the boundary.  Then, 
{\color{red} $$m_\mu := N_{1_\cC}^\mu$$ 
(where $1_{\cC}$ is the tensor unit object of $\cC$)  is  the multiplicity of $\mu$ in the Dirichlet Lagrangian algebra. 
A \emph{charge} is an anyon $\mu$ such that $m_\mu \geq  1$, }i.e. an anyon which is condensed on the reference boundary.  A \emph{fixed-charge duality} (FCD) is a symmetry of the bulk TQFT (a braided autoequivalence of the MTC) which preserves each charge. \cite{etingof2009fusioncategorieshomotopytheory, Barkeshli_2019, aksoy2025}.

Since we are interested in SPT phases, we can take the physical boundary condition to be topological as well, given by a Lagrangian algebra $\cA_{\mathbb{B}_Q}$. Then $Z_\mu := \dim V_\mu$ is the multiplicity of the anyon $\mu$ appearing in the physical boundary Lagrangian algebra:
$$\cA_{\mathbb{B}_Q} = \bigoplus \limits_{\mu \in {\rm Obj}(Z(\cC))} Z_\mu  \mu.$$

Given this setup, how a duality $\cD$  changes the symmetries has a natural interpretation in the symTFT picture: we can simply compare  $F(\mu)$ and  $F(\cD(\mu))$. Since all boundary lines come from applying $F$ to some bulk line $\mu$,  if $F(\mu) = F(\cD(\mu))$ for all $\mu$, we can say that the duality $\cD$ preserves the symmetries. We prove this is the case for FCDs in the following theorem:

 \vspace{3mm}
\textbf{Theorem}: \textit{Let  $\cD$ be a symmetry of the bulk TQFT $Z(\cC)$, and $F: Z(\cC) \ra \cC$ the forgetful functor. Then,  $F(\cD(\mu)) = F(\mu) $ for any bulk anyon $\mu$ of $Z(\cC)$ if and only if $\cD$ is an FCD.}
\vspace{3mm}

\textbf{Proof}: We place the 1+1d system on a torus. The  $a$-twisted sector partition function  (i.e. line operator $a$ is inserted along time) is given by:
 \begin{align}
     Z_a^1 = \sum_\mu \Tr_{W_a^\mu \otimes V_\mu} \mathds{1} = \sum_\mu \Tr_{W_a^\mu}  \mathds{1} \Tr_{V_\mu} \mathds{1}= \sum_\mu N_a^\mu Z_\mu
     \label{eq:Za1}
 \end{align} 
 where we used $\Tr_{V_\mu} \mathds{1} = \dim V_\mu =  Z_\mu$, and the sum is over anyons $\mu$ of $Z(\cC)$. \footnote{For convenience we took the physical boundary condition to be topological, but the relations \ref{eq:Za1} and \ref{eq:Z1a} --  generalized Fourier transforms -- between symmetry sectors and anyon sectors holds in general \cite{choi2024generalizedtubealgebrassymmetryresolved}. If the physical boundary condition is conformal and the boundary manifold is a torus, for example,  $Z_\mu$ will be a partition function which is not a constant but depends on the modular parameter of the boundary torus.}
On the other hand, if we insert the symmetry line $a$ along space, we get
\begin{align}
    Z_1^a = \sum_\mu \Tr_{W_1^\mu \otimes V_\mu} [\cL_a] 
    =  \sum_\mu (\Tr_{W_1^\mu} \cL_a) Z_\mu
    \label{eq:Z1a}
\end{align}
where we denote by $\cL_a$ the line operator corresponding to the simple object $a$ of $\cC$, and used the fact that $\cL_a$ is decoupled from the physical boundary.  
Note that unless $m_\mu  = \dim W_{1_\cC}^\mu \geq 1$, we get  no contribution from the corresponding $\mu$, as $W_1^\mu$ is empty. 
Hence we  write 
$$Z_1^a = \sum \limits_\mu B_{a, \mu} Z_{\mu} $$
where the important property of the coefficients 
$$B_{a, \mu} := (\Tr_{W_1^\mu} \cL_a)$$ 
is that they  are zero unless $\mu$ is a charge.

The partition functions $Z_a^1$ and $Z_1^a$ are related by the modular $S$-transformation:
\begin{align}
    Z_a^1  = S \cdot Z^a_1 = S \cdot \sum_\mu  B_{a, \mu} Z_\mu = \sum_{\mu, \nu} B_{a, \mu} S_{\mu \nu} Z_\nu
    \label{eq:S-transform}
\end{align} 
where $S_{\mu\nu}$ is the $S$-matrix of the bulk TQFT.

Now, we will prove that $\cD$ is an FCD $\implies$ $F(\cD(\mu)) = F(\mu)$ for all $\mu$.  Suppose that $\cD$ is an FCD. Then, $Z^a_1$ is invariant under $\cD$ since only charges $\mu$ contribute to the sum. Moreover, since any duality leaves $S$ invariant, and $Z_a^1$ is merely $S$-transformed $Z^a_1$, $Z_a^1$ is also invariant under $\cD$. 

Explicitly,  denoting $\mu' = \cD(\mu)$, we compute 
\begin{align}
    \cD \cdot Z_a^1  =  \cD \cdot S^\dagger \cdot Z_1^a \nonumber \\ 
    =     \sum_{\mu, \nu} B_{a, \mu} S^{\dagger}_{\mu, \nu}  Z_{\nu'} = \sum_{\mu', \nu'} B_{a, \mu'} S^{\dagger}_{\mu' \nu'} Z_{\nu'} \nonumber  \\ 
    = S^{\dagger} \cdot Z_1^a  = Z_a^1,
\end{align}
where we have used the fact that $B_{a, \mu} = B_{a, \mu'}$ and  $S_{\mu' \nu'} = S_{\mu \nu}.$
Applying the invariance of $Z_a^1$ to Eq. \eqref{eq:Za1}, we get
\begin{align}
N_a^{\cD(\mu)} = N_a^\mu
\end{align} for all $a$ and $\mu$. Recalling Eq. \eqref{eq:F}, this  shows that if $\cD$ is an FCD, then $F(\mu) = F(\cD(\mu))$ for all  simple objects $\mu$ of $Z(\cC)$. 

Now we prove the reverse direction, that $F(\cD(\mu)) = F(\mu)$ for all $\mu$ $\implies$ $\cD$ is an FCD. Suppose a duality $\cD$ satisfies $F(\cD(\mu)) = F(\mu)$ for all $\mu$. Then, reversing the above logic, $Z_a^1$ is invariant under $\cD$, which in turn means $Z_1^a$ is also invariant under $\cD$ since $Z_1^a$ is related to $Z_a^1$ via $S$-transformation.
Invariance of $Z_1^a$ under $\cD$ gives us, from Eq. \eqref{eq:Z1a} and the definition of $B_{a, \mu}$, 
\begin{align}
    B_{a, \mu} = B_{a, \cD(\mu)}.
    \label{eq:Bduality}
\end{align}
for all $a$ and $\mu$.

Now we invoke the following lemma:

\vspace{3mm}
\textbf{Lemma}: \textit{If $B_{a, \mu} = B_{a, \nu}$ for all simple objects $a$ of $\cC$ and charges $\mu, \nu$  of $Z(\cC)$, then $\mu = \nu$.
}
\vspace{3mm}

The proof is given in Appendix \ref{app:lemma}.
Applying the lemma to Eq. \eqref{eq:Bduality}, we see that $\cD(\mu) = \mu$ if $\mu$ is a charge (if $\mu$ is not a charge, $B_{a, \mu}$ is identically zero so the equation is trivially satisfied). Hence, we conclude that, if a duality $\cD$ satisfies  $F(\cD(\mu)) = F(\mu)$, then $\cD$ fixes charges. $\square$

\vspace{3mm}

While this theorem is a statement about TQFT/fusion categories, we know that dualities of the 1+1d $\cC$-symmetric lattice models  arise from symmetries of the 2+1d bulk symTFT \cite{moradi2023topological, Lootens_2023, Lootens_2024, Lootens_2025}. \footnote{  More precisely, the general construction provided in Refs. \cite{Lootens_2023, Lootens_2024, Lootens_2025} builds a lattice duality starting from a $\cD$-module functor where $\cD$ is some fusion category Morita equivalent to the symmetry category $\cC$. In principle, a symmetry of the bulk symTFT $Z(\cD) \simeq Z(\cC)$ gives rise to such a $\cD$-module functor.} In particular,  dualities which preserve the Dirichlet boundary condition correspond to FDQCs, which map SPT phases to SPT phases. \footnote{ For an explicit example on the lattice (for the grouplike $\ZZ_2 \times \ZZ_2$ cluster state) of the bulk symmetry giving rise to boundary duality/symmetric entangler, see  Sec. 5.2.1 of Ref. \cite{ji2025bulkexcitationsinvertiblephases}.} 
 The theorem says that, when the duality moreover preserves individual charges, it commutes with the symmetries, and given the fruitful connection between TQFTs and gapped lattice systems, it is reasonable to expect that this property also holds on the lattice level. 
This leads us to conjecture that, when  a symmetric entangler connecting the two SPT phases will exist if and only if there is an FCD connecting them. \footnote{While it is possible that such an FCD could only be realized by a locality-preserving unitary circuit that is not finite-depth, we  conjecture that for 1+1d it should be finite-depth, by analogy to the invertible symmetry case,  where quantum cellular automata (QCA) are only necessary for beyond-group cohomology phases where coupling to some spacetime structure beyond the symmetry defects become important. The example of Sec. \ref{sec:A4}, which is indeed finite-depth (i.e. a QCA of index zero), provides strong evidence.}

We conclude this section by noting that the conjecture is consistent with known examples of non-invertible SPT phases. First, 
$\Rep(D_8)$ has three SPT phases, which have been explicitly constructed on spin chains in  Ref. \cite{seifnashri2024cluster}. These phases are not connected by any FCD \cite{aksoy2025}, so we expect no symmetric entangler exists. This is consistent with the result of  Ref. \cite{seifnashri2024cluster}, which  showed that no symmetric entangler  connecting these SPT states exists.

Ref. \cite{Fechisin_2025} constructed generalized cluster states for $\cC = {\rm Vec}_G \times \Rep(G)$ symmetry, which belong to an SPT phase distinct from that of the product state. For non-abelian $G$, these provide a class of SPT states protected by non-invertible symmetries.  

The symTFT is given by $Z(\cC)  = D(G) \boxtimes D(G),$ where $D(G) \simeq Z({\rm Vec}_G)$. The   anyons of  $D(G)$ can be  written  as pairs $([g], \rho)$ where  $[g]$ is a conjugacy class of $G$ and $\rho$ is an irrep of the centralizer $C_G(g)$ of a representative $g$ of $[g]$. Generalizing the $G = S_3$ case described in Appendix I of Ref. \cite{Fechisin_2025},  
we see that in terms of  Lagrangian algebras, the product state phase corresponds to 
\begin{align}
    \cA_1 = \left( \bigoplus_{[g]} ([g], 1) \right) \boxtimes \left( \bigoplus_{R} ([e], R)\right)
\end{align}
(here, the first sum is over all conjugacy classes $[g]$ of $G$ and the second sum is over all irreps $R$ of $G$), 
  whereas the cluster state phase corresponds to 
  \begin{align}
      \cA_2 = \bigoplus_{\mu \in {\rm Obj}(D(G))} d_\mu \mu \otimes \bar{\mu}
  \end{align}
  where $d_\mu$ is the quantum dimension of an anyon $\mu$ of $D(G).$

   In $\cA_1$, each anyon appears with multiplicity $1$, whereas in $\cA_2$, there are  at least some anyons which appear with multiplicity $\geq 2$ (since for non-abelian $G$, $D(G)$ is a non-abelian MTC). Since an FSD can at most permute anyons, it is impossible to change the multiplicity, and thus there is not even an FSD connecting  the two phases. \emph{A fortiori}, there is no FCD connecting the $G \times \Rep(G)$ cluster state phase to the product state phase. Thus, we expect that a symmetric entangler does not exist. 
  
  Note that, while Ref. \cite{Fechisin_2025}
provides a unitary operator $U_\cC$ (in Eq. 23) which maps the product state to the cluster state, this operator actually does not commute with the symmetries when $G$ is non-abelian. Thus, their result is consistent with ours. 

These known examples are naturally all negative, as the existence of symmetric entanglers for non-invertible symmetries had not been recognized. In the following section, we construct the first positive example.

\section{Construction of symmetric entangler for \texorpdfstring{$\Rep(A_4)$}{} SPT phases}
\label{sec:A4}

There are  two SPT phases with $\Rep(A_4)$ symmetry, connected by an FCD \cite{aksoy2025}. Hence, according to our conjecture, we expect the entangler mapping between these two phases to commute with the symmetries. We will  construct two states belonging to the two SPT phases, and then  construct a symmetric entangler connected those states. 

We first fix some notation for the alternating group $A_4$, which is of order $12$.  We present $A_4$ with two generators $x$ and $a$ such that 
$$x^3 = a^2 = e,  \quad (xa)^3 = e.$$ 
We also define, for convenience, 
$$b \equiv xax^{-1}, \quad  c \equiv xbx^{-1}.$$
$\{ e, a, b, c \}$ form  the $\ZZ_2\times \ZZ_2$ subgroup of $A_4$: 
$$a^2= b^2 = c^2, c = ab.$$
There are four conjugacy classes, $[e], [a], [x],$ and $[x^2]$, of sizes $1, 3, 4,$ and $4$, respectively.

Recall that a $\Rep(G)$-symmetric phases are classified by module categories $\Rep^\psi(H)$  over $\Rep(G)$ \cite{etingof2016tensor}. Here, $\Rep^\psi(H)$ is the category of $\psi$-twisted projective representations of $H$, where $\psi \in H^2(H, \CC^\times)$ is the group $2$-cocycle for the projective representations. The SPT phases, which have a single simple object/vacuum, are given by module categories $\Rep^\psi(H)$ such that there is a unique $\psi$-twisted projective irrep. For $G = A_4$, we have two SPT phases, given by $\Rep(1)$ (where $1$ denotes the trivial subgroup) and $\Rep^\omega(\ZZ_2\times\ZZ_2)$, where $\omega$ is a nontrivial $2$-cocycle for $\ZZ_2 \times \ZZ_2$ (recall that $\ZZ_2\times\ZZ_2$ has a unique nontrivial projective irrep, of degree $2$). 

Following Ref. \cite{meng2025noninvertiblesptsonsiterealization}, we construct the SPT states as matrix product states (MPS) on the closed chain.  The  physical Hilbert  space on each site is given by 
$$\cH_i = \CC[G],$$
and the group elements $ k \in G$ give us basis states 
$$|k\rangle_i \in \cH_i.$$
The symmetries act as follows: denote by $\bf{1},  \omega, \omega^2, \pi$ the four irreps of $A_4$, where $\bf{1}, \omega, \omega^2$ are the $1$d irreps and $\pi$ is the $3$d. The $1$d irreps are defined by ($\bf{1}$ is the trivial irrep)
$$\omega(x) = e^{2\pi i /3}, \quad  \omega^2(x) = e^{ -2 \pi i /3} ,$$
with $\ZZ_2 \times \ZZ_2$  the kernel of $\omega$ and $\omega^2$; $\pi$ is faithful and an explicit form for  the generators is given by 
$$\pi(x) = \begin{pmatrix}
    0 & 1  & 0 \\
    0  & 0 & 1 \\
    1 & 0 & 0 
\end{pmatrix}, \quad \pi(a) = \begin{pmatrix}
    1 & 0 & 0 \\
    0 & -1 & 0 \\ 
    0& 0& -1
\end{pmatrix}.$$
Recall that a matrix product operator (MPO) with tensors $T^{k,l}$ (which are matrices -- endomorphisms of the bond space -- for fixed $k, l$) is defined, on the closed chain with $N$ sites, as \cite{Garre_Rubio_2023, meng2025noninvertiblesptsonsiterealization}
\begin{align}
    &O_T^N \nonumber \\ 
    = & \sum \limits_{ \{k_i,l_i\} } \Tr[T^{k_1, l_1}  \cdots T^{l_N, k_N}] |k_1,  \cdots , k_N\rangle \langle l_1, \cdots,  l_N|. \nonumber
    \end{align}
The symmetry operators $\cL_{\bf{1}}, \cL_\omega, \cL_{\omega^2}, \cL_\pi$ are defined as MPOs with tensors 
\begin{align}
    T_{\bf{1}}^{k,l} = \delta_{k,l},  \nonumber \\ 
    T_{\omega}^{k,l} = \delta_{k,l} \omega(k) ,\nonumber \\
    T_{\omega^2}^{k,l} = \delta_{k,l} \omega^2(k), \nonumber \\ 
    T_\pi^{k,l} = \delta_{k,l} \pi(k),
    \label{eq:MPOsym}
\end{align}
respectively, 
for $k, l \in A_4$
(see Ref. \cite{meng2025noninvertiblesptsonsiterealization} for a $\Rep(D_8)$ analogue).

Note that the MPOs for $1$d irreps have $1$d bond space and are trivial MPOs (they are a product of local operators), while the MPO corresponding to $\pi$ has $3$d bond space and is a nontrivial MPO.  It is easy to see that the action of these MPOs when acting on a general basis state 
$$|l_1, l_2, \cdots \rangle$$
only depends on the conjugacy class of the product  $l_1 l_2 \cdots$ of all group elements on each site. This conjugacy class corresponds precisely to the charge of such a basis state.

\subsubsection{SPT phase 1: product state}
We can construct an  SPT state corresponding to $\Rep(1)$ as  a product state 
\footnote{
While our construction represents this phase as a product state, we refrain from using the term ``trivial phase,'' as we cannot think of this phase as the unit with respect to stacking. }
\begin{align}
|\Psi_1\rangle  =  |e, e, e,\cdots \rangle     
\label{eq:MPS1}
\end{align}
where $e$ is the identity element. As an MPS, it is trivial: the bond space is $1$d. It is clear this is symmetric under the $\Rep(A_4)$ symmetry MPOs, as the MPO action only depends on the ``overall group element'' (the product of group elements over all sites). Note that 
$$\Tr[\pi(e)\pi(e)\cdots ]|e,e,\cdots\rangle = 3 |e,e \cdots \rangle.$$
For non-invertible symmetries, being ``symmetric'' means we get a  factor of the quantum dimension, which is $3$ for $\cL_\pi$, when acting on the symmetric state.

\subsubsection{SPT phase 2}
From here on, we will fix $G = A_4$ and $H = \ZZ_2 \times \ZZ_2 \subset G$. 

A  state corresponding to $\Rep^\omega(\ZZ_2\times \ZZ_2)$ can be constructed as follows. 
We define matrices
\begin{align}
 Q(e) = \mathds{1}, \quad   Q(a) = Z, \quad Q(b) = X, \quad Q(c) = iY
\end{align}
where $X, Y, Z$ are Pauli matrices acting on the bond space. We may think of $Q: H \ra \rm{GL}(2)$ as the  projective representation  of $H$. The SPT state is defined as an MPS with 2d bond space,   with tensors 
$$A^g = Q(g)$$
for $g \in H$ and zero otherwise. Explicitly, the MPS is \footnote{The form of this MPS is identical to that of the $\Rep(D_8)$ SPT phase MPSs constructed in Ref. \cite{meng2025noninvertiblesptsonsiterealization}. This is not surprising, since $\Rep^\omega(\ZZ_2 \times \ZZ_2)$ can also be thought of as a module category over $\Rep(D_8)$. However, both the Hilbert space and the symmetries  here are completely different compared to the $\Rep(D_8)$ case.}
\begin{align}
    |\Psi_2\rangle =\sum_{\{g_i \in H\} } \Tr [Q(g_1) Q(g_2) \cdots ] | g_1 ,g_2,  \cdots \rangle.
    \label{eq:MPS2}
\end{align}
Note that the product $Q(g_1)Q(g_2) \cdots$ is equal to some $Q(g)$ (up to an overall phase arising from projectiveness) for some $g \in H$ since $Q$ are projective representation matrices. $\Tr[Q(g)]$ is nonzero iff $g = e$, so this means the product of group elements 
$$g_1 g_2 g_3 \cdots =e$$
for any basis state 
$$|g_1, g_2, g_3 \cdots\rangle$$ contributing to $|\Psi_2\rangle$.
Since the $\Rep(A_4)$-symmetry action only depends on the overall group element, this state is indeed symmetric. 

We have verified that the two MPSs belong to inequivalent phases by computing their $L$-symbols (see Ref. \cite{Garre_Rubio_2023} for the definition of $L$-symbols). We provide the details  of the computation in Appendix \ref{app:Lsymbols}.


\subsubsection{Symmetric entangler as an MPU}
Now, we construct a symmetric entangler connecting the two SPT states.  First, $A_4$ has a unique projective irrep of degree $2$. We denote this again by $Q$, using the fact that the projective representation  $Q$ of $H$ defined before arise as a restriction of this  projective representation   to $H$ (up to some phase freedom). Explicitly, we choose 
$$Q(a) =Z$$
(as before) and 
$$Q(x) = \frac{1}{2}(- \mathds{1}  + i X  + iY + Z) = \frac{1}{2} \begin{pmatrix}
    -1+ i & 1+i\\
    -1 +i & -1-i
\end{pmatrix}.$$

Note that we can write a general element of $A_4$ in the form 
$$x^{n} g,$$
where $g \in H$, with well-defined $n$ mod $3$.  Then, consider an MPO tensor given in three ``blocks'' as:
\begin{align}
    T^{g,h} &=  \frac{1}{2}(s_e)_{g,h} Q(gh),  \nonumber \\
    T^{xg, xh} &= \frac{1}{2}(s_x)_{g,h} Q(x) Q(gh) , \nonumber \\ 
    T^{x^2g, x^2h} &= \frac{1}{2}(s_{x^2})_{g,h}(g,h) Q(x)^2 Q(gh) , 
\end{align}
and $T^{k,l} = 0$ if the $k$ and $l$ have different powers of $x$ involved; here, $g, h\in H$ 
(note that all elements of $H$ are order 2, so  $h^{-1} = h$). Concretely, we may think of this MPO tensor as a $12 \times 12$ matrix for the physical space (consisting of three diagonal blocks of $4 \times 4$ matrices), where each entry is itself a $2\times 2$ matrix for the bond space. Here, $s_e(g,h), s_x(g,h),$ and $ s_{x^2}(g,h)$ are some signs that depend on $g,h$, which are necessary to make the MPO unitary. 
Explicitly, we can  take 
\begin{align}
    (s_e)_{g,h} &=\begin{pmatrix}
    +  & + & - & + \\
    -  & + & + & + \\
    +  & + & - & + \\
    +  & - & - & - \\
\end{pmatrix}, \quad 
    (s_x)_{g,h} &=\begin{pmatrix}
    +  & - & - & + \\
    + & - & + & -\\
    +  & + & + & + \\
    +  & + & - & - \\
\end{pmatrix}
\label{eq:signs}
\end{align}
and $(s_{x^2})_{g,h} = s_e(g,h)$.

It is easily seen that this MPO, which we denote by $\cE$, connects the two SPT states (for any system size):
\begin{align}
    \cE|\Psi_1\rangle = \sum_{ \{g_i\} }\Tr[T^{g_1,e}T^{g_2,e} \cdots ] |g_1 ,g_2 \cdots \rangle \nonumber \\ 
    = \sum_{ \{g_i\} }\Tr[Q(g_1) Q(g_2) \cdots ] |g_1 ,g_2 \cdots \rangle = |\Psi_2\rangle 
\end{align}
(up to a possible overall sign arising from Eq. \ref{eq:signs}.)

This MPO is unitary for any system size $N$, since it satisfies the conditions of Theorem 1 of Ref. \cite{shukla2025simplegeneralequationmatrix} for any $N$ -- thus,  it is an MPU. An MPU is equivalent to a quantum cellular automaton (QCA), which in turn is equivalent to an FDQC if its index is zero \cite{Ignacio_Cirac_2017}. The index of our MPU is zero (see Definition IV.1 of Ref. \cite{Ignacio_Cirac_2017} for index of an MPU), so it provides an FDQC.

The MPO tensors satisfy 
$$T^{j,i *} = U T^{i,j} U^{-1}$$ with $U = i Y$, which means the tensors $T^{j, i *}$, which generate the Hermitian conjugate MPO, in fact generate the same MPO. Hence, 
$$ \cE^{-1} = \cE^\dagger =  \cE,$$ which means  $\cE$ is order $2$, and the two SPT phases form a torsor over $\ZZ_2$.

As we show in Appendix \ref{app:commutes},  this entangler commutes with the $\Rep(A_4)$ symmetry. Thus, putting everything together, our MPU is a globally symmetric FDQC, i.e. a symmetric entangler, connecting two inequivalent SPT states of $\Rep(A_4)$.

\section{Discussion}

We have argued, using topological holography, that an FCD will give rise to a symmetric entangler for non-invertible SPT phases, and  constructed an explicit example of a symmetric entangler for $\Rep(A_4)$ SPT phases.  This overturns the previous expectation, based on the lack of stacking structure for non-invertible symmetries, that such entanglers would not exist, and provides a microscopic basis for the classification framework proposed in Ref. \cite{aksoy2025}.

Given the close connection between symmetric entanglers and SPT-stacking in the invertible symmetry case, the symmetric entangler for non-invertible symmetries may be taken as implementing a stacking operation for non-invertible SPT phases, albeit only for those connected by FCDs. It remains to be seen whether a generally applicable notion of stacking can be defined for non-invertible symmetries.

 While we were guided in the quest to construct an explicit example by the conjecture in Sec. \ref{sec:FCD} that such a symmetric entangler should exist, the  construction of the explicit MPU was \emph{ad hoc}, and did not directly reference the FCD of the bulk symTFT. The example provides strong evidence for the conjecture that FCDs can be realized by symmetric entanglers, but does not prove it, nor is it yet clear how a symmetric entanglers can be constructed in general given an FCD. It would be interesting to see if a symmetric entangler can  be derived from a bulk FCD by restricting it to the boundary and show that the conjecture is true in general.

\appendix
\section{Proof of lemma in Sec. II}
\label{app:lemma}
Comparing Eqs. \eqref{eq:S-transform}  and \eqref{eq:Za1}, we see that $$B_{a,\mu} = \sum_\tau N^\tau_a S_{\tau \mu}.$$
We define the quantity 
\begin{align}
    M_{\mu\nu} := \sum_a B^*_{a,\mu} B_{a, \nu}.
\end{align}

Then, 
\begin{align}
    M_{\mu \nu} = \sum_a \sum_\tau \sum_\sigma  N^\tau_a S^*_{\tau \mu} N_a^\sigma S_{\sigma \nu}
\end{align}
since $N$ is real.

Now, we note that
\begin{align}
    \langle  (F(\tau), F(\sigma)) \rangle_{\cC} =\langle  \bigoplus_a N_a^\tau a, \bigoplus_b N_b^\sigma b \rangle_{\cC}   \nonumber \\
    = \sum_{a, b} N_a^\tau N_b^\sigma \delta_{a,b}  = \sum_a N_a^\tau N_a^\sigma.
    \label{eq:FF}
\end{align}
(Here, $\langle \cdot, \cdot \rangle_{\cC}$ denotes hom-space $\Hom_{\cC} (\cdot, \cdot )$ in category $\cC$.)

On the other hand, consider the induction functor $I: \cC \ra Z(\cC) $ which is adjoint to $F$, which satisfies 
$$I(1_{\cC}) = \bigoplus_\lambda m_\lambda \lambda ,$$ 
where $1_{\cC}$ denotes the tensor unit object of $\cC$) and $m_{\lambda}$ is the multiplicity of an anyon $\lambda$ in the reference boundary Lagrangian algebra (hence $m_\lambda \neq 0$ iff $\lambda$ is a charge).   
Then, Eq. \ref{eq:FF} is also equal to 
\begin{align}
    &\langle 1_{\cC}, F(\tau)^* \otimes F(\sigma) \rangle_{\cC} = \langle 1_{\cC}, F(\tau^* \otimes \sigma) \rangle_{\cC} \nonumber \\  
    &= \langle I(1), \tau^* \otimes \sigma \rangle _{Z(\cC)} 
    = \langle \bigoplus_\lambda m_\lambda \lambda , \tau^* \otimes \sigma \rangle_{Z(\cC)} \nonumber \\ 
   & = \sum_\lambda m_\lambda \langle \lambda \otimes \tau, \sigma \rangle_{Z(\cC)} = \sum_\lambda m_\lambda \cN_{\lambda \tau}^\sigma.
\end{align}
where $\cN$ are the fusion coefficients for $Z(\cC)$. 

Now, plugging the Verlinde formula 
$$\cN_{\lambda \tau}^\sigma = \sum_\kappa \frac{S_{\lambda\kappa} S_{\tau \kappa} S^*_{\sigma \kappa}  }{S_{0\kappa}} $$
into $M_{\mu \nu} =  \sum_\tau \sum_{\sigma} S^*_{\tau \mu} S_{\sigma \nu} \sum_\lambda m_\lambda \cN_{\lambda \tau}^\sigma, $
we get 
\begin{align}
    M_{\mu \nu} =  \sum_\tau \sum_{\sigma} S^*_{\tau \mu} S_{\sigma \nu} \sum_\lambda m_\lambda \sum_\kappa \frac{S_{\lambda\kappa} S_{\tau \kappa} S^*_{\sigma \kappa}  }{S_{0\kappa}} \nonumber \\ 
    = \sum_\kappa \sum_\tau   S^*_{\tau \mu} S_{\tau \kappa} \sum_\sigma S^*_{\sigma \kappa} S_{\sigma \nu} \sum_{\lambda} m_\lambda S_{\lambda \kappa} \frac{1}{S_{0 \kappa}}.
\end{align}
From the fact that $m_\lambda$ are coefficients of a Lagrangian algebra, and from $S^T = S$, we have 
$$S^T m = Sm = m.$$  Also, $S^\dagger = S^*$ so $S^* S = \mathds{1}$. Thus, we have 
\begin{align}
    M_{\mu \nu} = \sum_{\kappa} \delta_{\mu \kappa} \delta_{\kappa \nu} \frac{m_\kappa}{S_{0 \kappa}} = \delta_{\mu \nu} \frac{m_\mu}{S_{0\mu}}
\end{align}
which means $M_{\mu \nu}$ is a diagonal matrix, with the diagonals given by $\frac{m_\mu}{S_{0\mu}}.$ In particular, since $S_{0\mu} = \frac{d_{\mu}}{D} > 0$, $M_{\mu \mu} >0$ whenever $m_\mu > 0$, which means $\mu$ is a charge. 

Now suppose that $\mu$ is a charge -- i.e $m_\mu > 0$ -- and that $B_{a\mu} = B_{a\mu'}$. From the definition of $M$ we have 
$$M_{\mu \mu'} = \sum_a B^*_{a\mu} B_{a\mu'} = \sum_a B^*_{a \mu} B_{a \mu} = M_{\mu \mu}.$$

But if $\mu \neq \mu'$, $M_{\mu \mu'} = 0$ since $M$ is diagonal, and this means $M_{\mu \mu} = 0$ , but this is false according to the assumption that $m_\mu >0$. Thus we require $\mu = \mu'$. $\square$

 \section{The matrix product unitary commutes with \texorpdfstring{$\Rep(A_4)$}{} symmetry}
 \label{app:commutes}

We show that the $\Rep(A_4)$ entangler defined in Eq. (12) of the main text indeed commutes with the  $\Rep(A_4)$ symmetry.
 To this end, we consider how the MPU acts on a general basis state 
\begin{align}
    |x^{n_1} h_1, x^{n_2} h_2, \cdots \rangle 
\end{align}
of our Hilbert space. Since the symmetry action depends only on the conjugacy class of the overall group element $\prod_{i=1}^N x^{n_i} h_i$, it is sufficient to show that our MPU preserves the conjugacy class of the overall group element --  the symmetry action will then commute with the MPU.

Acting with the MPU gives us the state 
\begin{align}
\sum_{ \{g_i \}}\Tr[Q(x)^{n_1} Q(g_1h_1)Q(x)^{n_2} Q(g_2h_2) \cdots ]     |x^{n_1} g_1, x^{n_2} g_2, \cdots \rangle  
\end{align}
up to some signs from Eq. (13). 
Note that the block structure of the MPU tensor means that $n_i$ are preserved. This in turn preserves the overall factor of $x$ that appears in the overall group element. Thus, the conjugacy classes $[x]$ and $[x^2]$  are preserved. 

When the overall group element is $g \in H$, we need to preserve the classes $[e]$ and $[a]$ separately. To see that this is the case, we first note that $\Tr[Q(g)]$ vanishes for $g \neq e$ for  $g \in H$, and since $Q$ is a  projective representation, only those states with 
$$\prod_{i=1}^N x^{n_i}  g_i h_i = e$$
contribute. 

Now, we can commute all $x^{n_i}$ past all the elements (let's say, to the left); the $x$ factor then vanishes (since we are assuming the overall group element lives in $H$). We then have 
$$x^{n_1} h_1 x^{n_2} h_2 \cdots = h'_1 h'_2 \cdots $$
for some $h'_i \in H$ and  
$$x^{n_1} g_1 x^{n_2} g_2 \cdots = g'_1 g'_2 \cdots $$
for some $g'_i \in H$, since commuting $x^{n_i}$ past a $g_i$ or $h_i$ does not take it out of $H$. We also have  
$$x^{n_1} g_1 h_1 x^{n_2} g_2 h_2 \cdots = g'_1 h'_1 g'_2 h'_2\cdots,$$
which must equal $e$ for the trace to be nonvanishing.  Now, since  $g'_i, h'_i \in H$, they all  commute with each other, so we have 
$$(g'_1 g'_2 \cdots) (h'_1 h'_2\cdots) = e,$$
which in turn implies 
$$g'_1 g'_2 \cdots = h'_1 h'_2 \cdots$$
since all elements of $H$ are order $2$. This  means every basis state 
$$|x^{n_1} g_1, x^{n_2} g_2, \cdots \rangle $$
arising after acting with the MPU on a general basis state 
$$|x^{n_1} h_1, x^{n_2} h_2, \cdots \rangle$$
has the same overall group element as the latter state -- i.e. we preserve the charge. Hence the MPU commutes with the symmetry.

\section{\texorpdfstring{$L$}{}-symbols and  torus partition functions for \texorpdfstring{$\Rep(A_4)$}{} SPT phases}
\label{app:Lsymbols}

In this appendix, we will verify that the two $\Rep(A_4)$ SPT states constructed in the main text belong to distinct phases by computing their $L$-symbols, following the formalism of Refs. \cite{Garre_Rubio_2023, inamura202411dsptphasesfusion}. We will restrict our attention to $\Rep(G)$ SPT phases: hence the quantum dimensions are integers (equal to the dimension of the $G$-representation), and the MPS is injective. 

First, recall that, given MPO tensors $T_a^{l,m}$ for each $a \in \cC$ (with quantum dimension $d_a$) for some fusion category $\cC$, the fusion tensors $W_{ab}^{c;\mu}: \CC^{d_a} \otimes \CC^{d_b} \ra \CC^{d_c}$ are tensors which satisfy 
\begin{align}
 \sum_{m}  T_a^{l,m} T_b^{m,p}  = \sum_{c, \mu} ({W}_{ab}^{c; \mu})^{\dagger} T_c^{l,p} W_{ab}^{c ; \mu}  
\end{align}
for all $a, b,c \in \cC$ and $\mu = 1, ..., N_{ab}^c$. (Here and elsewhere, matrix multiplication over the bond space is implicit.) 

Given MPS tensors $A^l$ with bond space dimension $D$, we also have action tensors $V_a^i: \CC^{d_a} \otimes \CC^D \ra \CC^D$, which fractionalizes the action of the MPO $T_a^{l,m}$ on the MPS on the bond space:
\begin{align}
    \sum_lT_a^{m,l} A^l = \sum_i (V_a^{i})^{\dagger}   A^m  V_a^i
\end{align}
where $i = 1,...,d_a$. 

The $L$-symbols, which generalize the group $2$-cocycle for ordinary symmetries, can be computed from the fusion and action tensors  as
 \begin{align}
     (L_{ab})^{ij}_{c, k;\mu} = \frac{1}{D } \Tr[  V_a^i(\mathds{1}_{d_a} \otimes V_b^j)  ((W^{c, \mu}_{ab})^{\dagger} \otimes \mathds{1}_D) (V_c^k)^{\dagger}]
 \end{align}
 and the inverse $L$-symbols as 
 \begin{align}
     (\bar{L}_{ab})_{c , \mu}^{ij} = \frac{1}{D } \Tr[    V_c^k (W^{c, \mu}_{ab} \otimes \mathds{1}_D)     (\mathds{1}_{d_a} \otimes (V_b^j)^\dagger) (V_a^i)^\dagger ]
 \end{align}
 where $D$ is the bond space dimension of the MPS. 

The $L$-symbols, like the  $2$-cocycle, are not gauge-invariant quantities. To construct a   gauge-invariant quantity out of $L$-symbols, we can look at the torus partition functions of the 1+1d $\cC$-symmetric TQFT corresponding to the $L$-symbols. With an insertion of lines $a, b$ along the two cycles of the torus, and with a specification of an ``internal'' line $c$ resolving the crossing of $a$ and $b$, and with a specification of $\mu, \nu$ on each trivalent vertex (there is still gauge freedom in this $\mu, \nu$ space), we get \cite{inamura202411dsptphasesfusion}
\begin{align}
    (Z_{ab,c})_{\mu \nu} = \sum_{i,j,k} (L_{ab})_{c; k, \mu}^{ij} (\bar{L}_{ba})_{c;k, \mu }^{ji}.
\end{align}

Now, we move on to concrete computation for the two $\Rep(A_4)$-SPT states. Since the invertible symmetry generators act completely trivially on the MPS tensors for both states, we focus only on the non-invertible symmetry generator $\pi$. Recall that the fusion rules for $\pi$ involve fusion multiplicities:
$$\pi \otimes \pi = \mathbf{1} \oplus \mathbf{\omega} \oplus \mathbf{\omega}^2 \oplus 2 \pi. $$
Hence, we have a nontrivial index $\mu = 1, 2$ for the fusion channel ${\pi \pi} \ra \pi$. 
Note  that the MPO tensor for the symmetry generator $\pi \in \Rep(A_4)$ satisfies 
$$\sum_{k} T_\pi^{g,k} T_\pi^{k,h} = \sum_k \pi(g) \delta_{g,k}  \otimes \pi(k) \delta_{k,h} = \pi(g) \otimes \pi(g) \delta_{g,h}.$$
Then, the fusion tensors $W_{\pi \pi}^{c, \mu}: \CC^3 \otimes \CC^3 \ra \CC^{d_c}$ need to satisfy 
    \begin{align}\sum_\mu \sum_c (W_{\pi\pi}^{c, \mu})^{\dagger} \rho_c (g) W_{\pi\pi}^{c, \mu} = \pi(g) \otimes \pi(g)
\end{align}
for all $g \in A_4$
(where $\rho_c$ is a representation of $A_4$ corresponding to the simple object $c \in \Rep(A_4)$). They can  be  represented as $9 \times d_c$ matrices and chosen as  
\begin{align}
W_{\pi \pi}^{\mathbf{1}} &= \frac{1}{\sqrt{3}} \begin{pmatrix}
    1 & 0 & 0 & 0 & 1 & 0 & 0 & 0 & 1
\end{pmatrix},\nonumber \\ 
W_{\pi \pi}^{\mathbf{\omega}} &= e^{2 \pi i/ 3} W_{\pi \pi}^{\mathbf{1}}, \nonumber \\ 
W_{\pi \pi}^{\mathbf{\omega}^2} &=e^{ -2\pi i /3} W_{\pi \pi}^{\mathbf{1}},\nonumber \\ 
    W_{\pi \pi}^{\pi; 1} &=\frac{1}{\sqrt{2}} \begin{pmatrix}
        0 & 0 & 0 & 0 & 0 &-1 &0 &1 &0 \\
        0 & 0 & 1& 0& 0  &0 & -1 &0 &0 \\
        0 & -1 &0 & 1& 0 &0 &0 &0 &0 
    \end{pmatrix}, \nonumber \\ 
      W_{\pi \pi}^{\pi; 2} &=\frac{1}{\sqrt{2}} \begin{pmatrix}
        0 & 0 & 0 & 0 & 0 &1 &0 &1 &0 \\
        0 & 0 & 1& 0& 0  &0 &1 &0 &0 \\
        0 & 1 &0 & 1& 0 &0 &0 &0 &0 
    \end{pmatrix}
\end{align}

For  SPT phase 1, given by the product state, the action tensors for $\pi$ are maps $V_\pi^i: \CC^3 \ra \CC $, and can  simply be chosen as 
\begin{align}
    V_\pi^1 = \begin{pmatrix}
        1 & 0 & 0 
    \end{pmatrix}, \nonumber \\ 
        V_\pi^2= \begin{pmatrix}
        0 & 1 & 0 
    \end{pmatrix} , \nonumber \\
        V_\pi^3 = \begin{pmatrix}
        0 & 0 & 1 
    \end{pmatrix}.
\end{align}

From these and the fusion tensors, we can compute the $L$-symbols and the torus partition functions. Representing the torus partition function with insertion of lines $a = b = c = \pi$, and representing it as a $2 \times 2$ matrix in the $\mu, \nu$ indices, we get 
\begin{align}
  (  Z_{\pi\pi, \pi})_{\mu \nu} = \begin{pmatrix}
      -3  & 0 \\ 0 & 3
  \end{pmatrix}
\end{align}

For SPT phase 2, the action tensors $V_\pi^i : \CC^3 \otimes \CC^2 \ra \CC^2$, which we represent as $6 \times 2$ matrices, can be chosen as
\begin{align}
      V_\pi^1 = \begin{pmatrix}
        1 & 0 & 0 & 0 & 0 &0 \\
        0 & - 1 &  0 & 0 & 0 & 0
        \end{pmatrix}, \nonumber \\ 
         V_\pi^2 = \begin{pmatrix}
        0 & 0 & 0 & 1 & 0 &0 \\
        0 & 0 &  -1 & 0 & 0 & 0
        \end{pmatrix}, \nonumber \\
         V_\pi^3 = \begin{pmatrix}
        0 & 0 & 0 & 0 & 0 &1 \\
        0 & 0  &  0 & 0 & 1 & 0
        \end{pmatrix}.
\end{align}
The torus partition function is 
\begin{align}
  (  Z_{\pi\pi, \pi})_{\mu \nu} = \begin{pmatrix}
      3  & 0 \\ 0 & -3
  \end{pmatrix}.
\end{align}
Hence we see that the partition functions $(Z_{\pi\pi, \pi})_{\mu \nu}$ for the two states differ by a sign. We still have the freedom to perform a basis transformation on the 2d space for the indices  $\mu ,\nu$, but this cannot get rid of the sign difference (since this basis transformation applies to both partition functions simultaneously).  Thus, the two states belong to two different SPT phases. 

We note an interesting fact about these $L$-symbols. In many cases, nontrivial $L$-symbols signal  a projective realization of the symmetry group/algebra, i.e. the a pair of symmetry generators which commute may only commute up to a phase when we look at the fractionalized action on the bond space. This happens not only for grouplike symmetries, but also for $\Rep(D_8)$  \cite{seifnashri2024cluster, meng2025noninvertiblesptsonsiterealization}. For the two $\Rep(A_4)$ SPT phases, however, only one symmetry generator, $\cL_\pi$, acts nontrivially, so there cannot be a projective realization of the symmetry algebra. Nevertheless,  the internal structure of how $\cL_\pi$ action fractionalizes on the bond space  is complex enough  to lead to inequivalent $L$-symbols for the  two phases.

\bibliography{ref}

\end{document}